# CoverM: Read alignment statistics for metagenomics

Samuel T. N. Aroney[1], Rhys J. P. Newell[1], Jakob N. Nissen[2], Antonio Pedro Camargo[3,4], Gene W. Tyson[1], Ben J. Woodcroft[1]*

[1]Centre for Microbiome Research, School of Biomedical Sciences, Queensland University of Technology (QUT), Translational Research Institute, Woolloongabba, Australia
[2]The Novo Nordisk Foundation Center for Protein Research, University of Copenhagen, Blegdamsvej 3A, Copenhagen, 2200, Denmark
[3]Departamento de Genética e Evolução, Instituto de Biologia, Universidade Estadual de Campinas, Campinas, São Paulo, Brazil
[4]DOE Joint Genome Institute, Lawrence Berkeley National Laboratory, Berkeley, California, USA

* To whom correspondence should be addressed.

# Abstract

## Summary

Genome-centric analysis of metagenomic samples is a powerful method for understanding the function of microbial communities. Calculating read coverage is a central part of analysis, enabling differential coverage binning for recovery of genomes and estimation of microbial community composition. Coverage is determined by processing read alignments to reference sequences of either contigs or genomes. Per-reference coverage is typically calculated in an ad-hoc manner, with each software package providing its own implementation and specific definition of coverage. Here we present a unified software package CoverM which calculates several coverage statistics for contigs and genomes in an ergonomic and flexible manner. It uses 'Mosdepth arrays' for computational efficiency and avoids unnecessary I/O overhead by calculating coverage statistics from streamed read alignment results.

## Availability and Implementation

CoverM is free software available at https://github.com/wwood/coverm. CoverM is implemented in Rust, with Python (https://github.com/apcamargo/pycoverm) and Julia (https://github.com/JuliaBinaryWrappers/CoverM_jll.jl) interfaces.

## Introduction

Microbial communities significantly impact the health of our planet and ourselves personally. Through the advent of genome-centric metagenomics, it has become possible to recover complete or near-complete genomes from these populations through deep sequencing of community DNA. Piecing together sequencing reads into metagenome assembled genomes (MAGs) has become increasingly common and high-throughput, with recovery of tens to hundreds of thousands of genomes from a diverse array of environments (Pasolli *et al.*, 2019; Almeida *et al.*, 2021; Nishimura and Yoshizawa, 2022; Schmidt *et al.*, 2024).

A key step in MAG recovery is binning, where assembled contigs are partitioned into groups representing putative genomes. Commonly, contig abundances across samples are calculated based on alignments of reads to contigs. Contigs with similar abundances across many samples are likely to be from the same genome, enabling improved, sequence-agnostic genome recovery (differential coverage binning) (Albertsen *et al.*, 2013). However, the specific algorithms metagenomic binning tools use to derive coverage statistics from read alignments (BAM files (Li et al., 2009)) vary substantially. For instance, some use the number of aligned reads divided by contig length (Wu *et al.*, 2014; Pan *et al.*, 2022, 2023; Nissen *et al.*, 2021). Others calculate a coverage based on the number of reads aligned to each reference position, taking the mean or trimmed mean across all bases in a contig (Kang *et al.*, 2015; Imelfort *et al.*, 2014). Still others find approximate coverage using a k-mer counting approach eschewing the alignment step altogether for faster calculation (Shaw and Yu, 2024).

The backend data structure used for coverage calculation can also vary. Notably, 'Mosdepth arrays' were shown to be two times faster than naive methods for exact calculation of mean coverage (Pedersen and Quinlan, 2018). Mosdepth arrays record the change in the number of mapped segments at each position, such that coverage at a given position can be calculated by summing all preceding values.

Recovered MAGs can be used to estimate the community composition by aligning reads to their contigs and calculating coverage statistics across the genome. For example, calculating the mean coverage for a genome is achieved by summing the number of aligning bases across all contigs in a genome and dividing by their total length. Researchers may also opt to apply filters to prevent genomes being deemed present (i.e. having non-zero coverage) on the basis of a small number of positions having aligned reads (Castro *et al.*, 2018).

Despite the differences between per-contig and per-genome coverage statistics, they are conceptually similar enough to motivate a single tool that calculates both. Here we present CoverM, which uses Mosdepth arrays to efficiently calculate a number of per-contig and per-genome coverage statistics.

## Software Implementation

CoverM is a software tool for calculating coverage statistics of contigs or genomes with reads derived from metagenomes or microbial isolates (**Figure 1**). It is written in the Rust programming language, relying on independent tools for genome dereplication (Aroney *et al.*, 2024) and read alignment (Sahlin, 2022; Li, 2018, 2013). It makes substantial use of rust-htslib (Köster, 2016), which provides Rust bindings to HTSlib (Bonfield *et al.*, 2021) for parsing intermediate alignment (BAM) files.

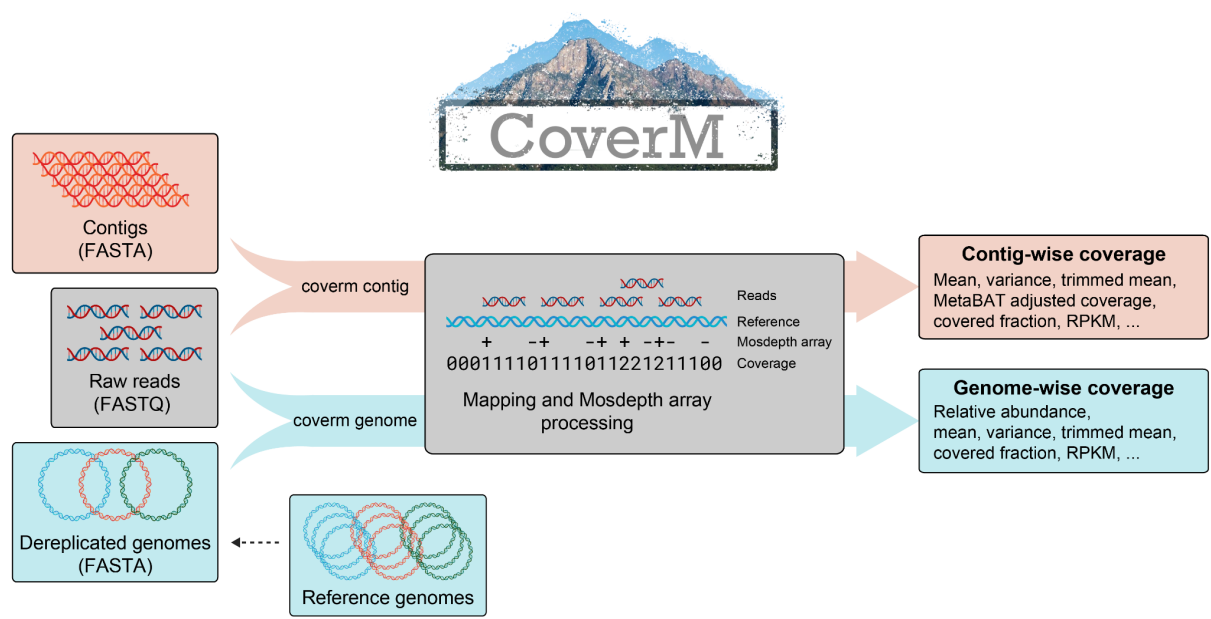

*Figure 1. CoverM operation.* In 'contig' mode, reads are aligned to contigs, alignments are sorted by reference position, a 'Mosdepth array' is constructed and then coverage statistics are calculated for each contig. In 'genome' mode, a Mosdepth array is similarly constructed for each contig in the genome, and then coverage statistics are reported for each genome.

CoverM can be compiled locally using the standard cargo method in the Rust ecosystem, by downloading the source code from the repository directly or by downloading a version statically compiled with musl libc from https://github.com/wwood/coverm/releases. Apart from the standard command line version, Python and Julia interfaces to CoverM are also available.

The inputs to CoverM are metagenomic reads (or precalculated BAM files) and reference sequences: either contigs or genomes. Since aligning against multiple near-identical reference genomes can distort read alignments, CoverM uses Galah (Aroney *et al.*, 2024) to dereplicate input genomes, interpreting ANI calculated by skani (Shaw and Yu, 2023), FastANI (Jain *et al.*, 2018) or Dashing (Baker and Langmead, 2019).

Next, the reads are aligned to the reference sequences using Strobealign (Sahlin, 2022), Minimap2 (Li, 2018) or BWA (Li, 2013). Spurious alignments can be filtered on the basis of identity, alignment length, or percent of the read aligned. Paired reads can be filtered independently or together with their mate. The filtered alignments are then processed using Mosdepth arrays to calculate a range of coverage statistics (**Table 1**). The coverage of a genome is aggregated by considering all bases in the genome together. As an additional filtering step, to account for off-target alignment in conserved regions, by default CoverM requires that at least 10% of a genome's length is covered by at least one read before assigning non-zero coverage.

| Table 1: Overview of calculation methods available in CoverM. The "reference" referred to in the description is either a contig or a set of contigs constituting a genome. The "ends of contigs" refers to the initial and final 75 bp of the contigs as per MetaBAT (Kang *et al.*, 2015). ||
|---|---|
| **Metric** | **Description** |
| mean (default contig mode) | Average coverage of each base, after excluding the ends of contigs. For genome mode, the average of contigs is calculated, weighting each contig by its length. |
| relative abundance (default genome mode) | Per-genome community profile percentage. See **Equation 1**. Genome mode only. |
| variance | Variance of each base's coverage, after excluding the ends of contigs. For genome mode, the variance is of all bases in the genome. |
| MetaBAT adjusted coverage | Identical output to the MetaBAT (Kang *et al.*, 2015) coverage calculation tool. |
| trimmed mean | Mean coverage after removing the most covered and least covered bases (default 5%). For genome mode, all bases in the genome are considered together. |
| covered fraction | Fraction of bases covered by at least one read, after excluding the ends of contigs. For genome mode, all bases in the genome are considered together. |
| covered bases | Number of bases covered by at least one read, after excluding the ends of contigs. For genome mode, all bases in the genome are considered together. |
| length | Length of each reference in base pairs. |
| count | Number of reads aligned. |
| reads per base | Number of reads aligned divided by reference length. |
| RPKM | Reads aligned to reference per kilobase per million reads aligned (Mortazavi *et al.*, 2008). |
| TPM | Reads aligned to reference per million reads aligned (Li *et al.*, 2010). |

The default per-genome coverage statistic is relative abundance of each genome (**Equation 1**). The relative abundance of a genome is an estimate of the proportion of the cells in a given sample belonging to the species with that genome (Sun *et al.*, 2021). Assuming no bias in DNA extraction, sequencing or read alignment, the ratio of genome copies in a sample is equal to the ratio of their mean coverages. One complication is that in many metagenomes the reference database is incomplete, so many reads remain unaligned. To account for this, the fraction of reads that are unaligned is used to scale the relative abundances of each genome.

$$relative\ abundance_i = \frac{mean\ coverage_i}{\sum_j mean\ coverage_j} \times \frac{aligned\ reads}{total\ reads} \times 100\% \qquad (eq.\ 1)$$

In **Equation 1**, $relative\ abundance_i$ is the estimated relative abundance of the lineage as a percentage, $mean\ coverage_i$ is the calculated mean coverage of that lineage, $\sum_j mean\ coverage_j$ is the sum of the mean coverage of all genomes, $aligned\ reads$ is the number of reads aligning to any genome, and $total\ reads$ is the total number of reads in the sequencing library. Implicitly, this formula assumes that (1) the average genome size of community members is the same between genomes in the reference set and those that are missing, and (2) species with reference genomes are roughly complete and uncontaminated.

## Conclusion

CoverM provides an efficient calculation method for several coverage metrics applicable to contig-based (metagenomic binning) and genome-based (community profiling) tasks. The implementation is flexible, with >50 arguments and options for controlling input/output formats, alignment, filtering, dereplication and coverage metric.

## Funding

This work was supported by the EMERGE National Science Foundation (NSF) Biology Integration Institute [grant number #2022070]; Genomic Science Program of the United States Department of Energy (DOE) Office of Biological and Environmental Research (BER) [grants numbers DE-SC0004632, DE-SC0010580 and DE-SC0016440]; Australian Research Council [grant numbers #DE160100248 to BW, #FT210100521 to BW]; and the São Paulo Research Foundation (FAPESP) [process number #2018/04240-0 to APC]. The work conducted by the US Department of Energy Joint Genome Institute (https://ror.org/04xm1d337) is supported by the US Department of Energy Office of Science user facilities, operated under contract no. DE-AC02-05CH11231.

## Reference list

Albertsen,M. *et al.* (2013) Genome sequences of rare, uncultured bacteria obtained by differential coverage binning of multiple metagenomes. *Nat. Biotechnol.*, **31**, 533–538.
Almeida,A. *et al.* (2021) A unified catalog of 204,938 reference genomes from the human gut microbiome. *Nat. Biotechnol.*, **39**, 105–114.
Aroney,S.T.N. *et al.* (2024) Galah: More scalable dereplication for metagenome assembled genomes.

Baker,D.N. and Langmead,B. (2019) Dashing: fast and accurate genomic distances with HyperLogLog. *Genome Biol.*, **20**, 265.
Bonfield,J.K. *et al.* (2021) HTSlib: C library for reading/writing high-throughput sequencing data. *GigaScience*, **10**, giab007.
Castro,J.C. *et al.* (2018) imGLAD: accurate detection and quantification of target organisms in metagenomes. *PeerJ*, **6**, e5882.
Imelfort,M. *et al.* (2014) GroopM: an automated tool for the recovery of population genomes from related metagenomes. *PeerJ*, **2**.
Jain,C. *et al.* (2018) High throughput ANI analysis of 90K prokaryotic genomes reveals clear species boundaries. *Nat. Commun.*, **9**, 5114.
Kang,D.D. *et al.* (2015) MetaBAT, an efficient tool for accurately reconstructing single genomes from complex microbial communities. *PeerJ*, **3**, e1165.
Köster,J. (2016) Rust-Bio: a fast and safe bioinformatics library. *Bioinformatics*, **32**, 444–446.
Li,B. *et al.* (2010) RNA-Seq gene expression estimation with read mapping uncertainty. *Bioinformatics*, **26**, 493–500.
Li,H. (2013) Aligning sequence reads, clone sequences and assembly contigs with BWA-MEM. *ArXiv13033997 Q-Bio*.
Li,H. (2018) Minimap2: pairwise alignment for nucleotide sequences. *Bioinformatics*, **34**, 3094–3100.
Mortazavi,A. *et al.* (2008) Mapping and quantifying mammalian transcriptomes by RNA-Seq. *Nat. Methods*, **5**, 621–628.
Nishimura,Y. and Yoshizawa,S. (2022) The OceanDNA MAG catalog contains over 50,000 prokaryotic genomes originated from various marine environments. *Sci. Data*, **9**, 305.
Nissen,J.N. *et al.* (2021) Improved metagenome binning and assembly using deep variational autoencoders. *Nat. Biotechnol.*, **39**, 555–560.
Pan,S. *et al.* (2022) A deep siamese neural network improves metagenome-assembled genomes in microbiome datasets across different environments. *Nat. Commun.*, **13**, 2326.
Pan,S. *et al.* (2023) SemiBin2: self-supervised contrastive learning leads to better MAGs for short- and long-read sequencing. *Bioinformatics*, **39**, i21–i29.
Pasolli,E. *et al.* (2019) Extensive Unexplored Human Microbiome Diversity Revealed by Over 150,000 Genomes from Metagenomes Spanning Age, Geography, and Lifestyle. *Cell*, **176**, 649-662.e20.
Pedersen,B.S. and Quinlan,A.R. (2018) Mosdepth: quick coverage calculation for genomes and exomes. *Bioinformatics*, **34**, 867–868.
Sahlin,K. (2022) Strobealign: flexible seed size enables ultra-fast and accurate read alignment. *Genome Biol.*, **23**, 260.
Schmidt,T.S.B. *et al.* (2024) SPIRE: a Searchable, Planetary-scale mIcrobiome REsource. *Nucleic Acids Res.*, gkad943.
Shaw,J. and Yu,Y.W. (2024) Fairy: fast approximate coverage for multi-sample metagenomic binning. *Microbiome*, **12**, 151.
Shaw,J. and Yu,Y.W. (2023) Fast and robust metagenomic sequence comparison through sparse chaining with skani. *Nat. Methods*, **20**, 1661–1665.
Sun,Z. *et al.* (2021) Challenges in benchmarking metagenomic profilers. *Nat. Methods*, **18**, 618–626.
Wu,Y.-W. *et al.* (2014) MaxBin: an automated binning method to recover individual genomes from metagenomes using an expectation-maximization algorithm. *Microbiome*, **2**, 26.